\documentclass[aps,prl,reprint,superscriptaddress]{revtex4-1}
\usepackage{graphicx}
\usepackage{dcolumn}
\usepackage{bm} 
\usepackage{braket} 
\usepackage{xcolor}

\begin{document}

\title{Enhanced charge density wave with mobile superconducting vortices in La$_{1.885}$Sr$_{0.115}$CuO$_4$}

\author{J.-J. Wen}
\email[]{jwen11@stanford.edu}
\affiliation{Stanford Institute for Materials and Energy Sciences, SLAC National Accelerator Laboratory, 2575 Sand Hill Road, Menlo Park, CA 94025, USA}
\author{W. He}
\affiliation{Stanford Institute for Materials and Energy Sciences, SLAC National Accelerator Laboratory, 2575 Sand Hill Road, Menlo Park, CA 94025, USA}
\affiliation{Department of Materials Science and Engineering, Stanford University, Stanford, CA 94305, USA}
\author{H. Jang}
\affiliation{Stanford Synchrotron Radiation Lightsource, SLAC National Accelerator Laboratory, Menlo Park, CA 94025, USA}
\affiliation{PAL-XFEL, Pohang Accelerator Laboratory, Gyeongbuk 37673, South Korea}
\author{H. Nojiri}
\affiliation{Institute for Materials Research, Tohoku University, Katahira 2-1-1, Sendai, 980-8577, Japan}
\author{S. Matsuzawa}
\affiliation{Institute for Materials Research, Tohoku University, Katahira 2-1-1, Sendai, 980-8577, Japan}
\author{S. Song}
\affiliation{Linac Coherent Light Source, SLAC National Accelerator Laboratory, Menlo Park, CA 94025, USA}
\author{M. Chollet}
\affiliation{Linac Coherent Light Source, SLAC National Accelerator Laboratory, Menlo Park, CA 94025, USA}
\author{D. Zhu}
\affiliation{Linac Coherent Light Source, SLAC National Accelerator Laboratory, Menlo Park, CA 94025, USA}
\author{Y.-J. Liu}
\affiliation{Stanford Synchrotron Radiation Lightsource, SLAC National Accelerator Laboratory, Menlo Park, CA 94025, USA}
\author{M. Fujita}
\affiliation{Institute for Materials Research, Tohoku University, Katahira 2-1-1, Sendai, 980-8577, Japan}
\author{J. M. Jiang}
\affiliation{Stanford Institute for Materials and Energy Sciences, SLAC National Accelerator Laboratory, 2575 Sand Hill Road, Menlo Park, CA 94025, USA}
\affiliation{Department of Applied Physics, Stanford University, Stanford, CA 94305, USA}
\author{C. R. Rotundu}
\affiliation{Stanford Institute for Materials and Energy Sciences, SLAC National Accelerator Laboratory, 2575 Sand Hill Road, Menlo Park, CA 94025, USA}
\author{C.-C. Kao}
\affiliation{SLAC National Accelerator Laboratory, Menlo Park, CA 94025, USA}
\author{H.-C. Jiang}
\affiliation{Stanford Institute for Materials and Energy Sciences, SLAC National Accelerator Laboratory, 2575 Sand Hill Road, Menlo Park, CA 94025, USA}
\author{J.-S. Lee}
\email[]{jslee@slac.stanford.edu}
\affiliation{Stanford Synchrotron Radiation Lightsource, SLAC National Accelerator Laboratory, Menlo Park, CA 94025, USA}
\author{Y. S. Lee}
\email[]{youngsl@stanford.edu}
\affiliation{Stanford Institute for Materials and Energy Sciences, SLAC National Accelerator Laboratory, 2575 Sand Hill Road, Menlo Park, CA 94025, USA}
\affiliation{Department of Applied Physics, Stanford University, Stanford, CA 94305, USA}


\maketitle

\noindent{{\bf Abstract}}

Superconductivity in the cuprates is found to be intertwined with charge and spin density waves \cite{Keimer2015,tranquada_review}. Determining the interactions between the different types of order is crucial for understanding these important materials \cite{RevModPhys.87.457}. Here, we elucidate the role of the charge density wave (CDW) in the prototypical cuprate La$_{1.885}$Sr$_{0.115}$CuO$_4$, by studying the effects of large magnetic fields ($H$) up to 24 Tesla. At low temperatures ($T$), the observed CDW peaks reveal two distinct regions in the material: a majority phase with short-range CDW coexisting with superconductivity, and a minority phase with longer-range CDW coexisting with static spin density wave (SDW). With increasing magnetic field, the CDW first grows smoothly in a manner similar to the SDW. However, at high fields we discover a sudden increase in the CDW amplitude upon entering the vortex-liquid state. Our results signify strong coupling of the CDW to mobile superconducting vortices and link enhanced CDW amplitude with local superconducting pairing across the $H-T$ phase diagram.\\

\noindent{{\bf Main}}

High-$T_c$ cuprates are the prominent example of a strongly correlated electronic system, featuring a rich phase diagram marked by novel types of order \cite{Keimer2015}. Tremendous effort has been devoted to studying these orders to understand the unconventional normal state and high-$T_c$ superconductivity \cite{RevModPhys.87.457}. It is recognized that such complex systems are susceptible to electronic inhomogeneity \cite{Dagotto257}, which may arise intrinsically due to electronic interactions \cite{EMERY1993597} or for extrinsic reasons, such as chemical disorder \cite{MARTIN200146}. Indeed, nanoscale spatial variations in electronic properties of the cuprates have been observed and may be relevant to their physics  \cite{pan2001,KRESIN2006231}. This issue is manifest in underdoped La$_2$CuO$_4$-based cuprates such as La$_{2-x}$Sr$_x$CuO$_4$ (LSCO), where long-range SDW (with correlation length of hundreds of unit cells) coincides with bulk superconductivity \cite{tranquada_review}. Consensus has yet to be reached regarding whether these orders coexist uniformly, or exist within distinct regions. Both scenarios have been suggested in theoretical studies \cite{Zhang1089,Kivelson11903,PhysRevLett.87.067202}, while it has been difficult to determine experimentally \cite{PhysRevB.66.014524,Mohottala2006,PhysRevB.78.104525}.

Recent observations of CDW order across cuprate families shed new perspective on the interplay between density wave orders and superconductivity \cite{Keimer2015,tranquada_review,RevModPhys.87.457}. When SDW order is absent, such as in Y-based cuprates (YBCO) around $1/8$ doping, clear competition between CDW and superconductivity has been observed \cite{Ghiringhelli821,Chang2012}. When both SDW and CDW are present in the La$_2$CuO$_4$-based cuprates, they appear to form a spin-charge stripe order pattern, indicated by nearly commensurate wave vectors ($q_{\rm cdw} \sim 2q_{\rm sdw}$) \cite{tranquada_review}. There are theoretical and experimental evidences of mutual cooperation between CDW and SDW order parameters when such commensuration is satisfied \cite{PhysRevB.57.1422,Wen2019}. Here the interaction between density waves and superconductivity appears more than simple competition. For example, the SDW onset temperature ($T_{\rm sdw}$) in LSCO is strikingly similar to the superconducting $T_c$ \cite{tranquada_review}, and a putative two-dimensional superconductivity called pair density wave has been associated with the stripe order \cite{RevModPhys.87.457,PhysRevLett.126.167001}. To elucidate the intrinsic behavior of these orders and distill universal features among different cuprates, it is important to carefully interpret observations in light of the inhomogeneity.

X-ray scattering combined with high magnetic fields provides a unique window into the nature of the coexistence of CDW, SDW, and superconductivity. If there is inhomogeneity on length scales larger than the CDW correlation length (typically tens of unit cells \cite{Ghiringhelli821,Chang2012,PhysRevB.89.224513,PhysRevB.90.100510,Wen2019,christensen2014}), it should be manifest in the evolution of the CDW peaks when the strength of the various orders is tuned by $T$ and $H$. We investigate La$_{1.885}$Sr$_{0.115}$CuO$_4$ where the stripe order is robust and of comparable strength to superconductivity ($T_{\rm sdw}\sim T_c$) \cite{PhysRevB.59.6517}. Our state-of-the-art x-ray free electron laser measurements (see Methods for details) reveal clear signatures of two types of CDW orders in the same sample, distinguished by different correlation lengths and distinct $T$- and $H$- dependences. Most importantly, we further uncover a strong connection between the short-range CDW correlations and mobile superconducting vortices, which bears important implications regarding the nature of the superconducting transition in the cuprates.

\begin{figure}[t]
 \includegraphics[scale=0.53]{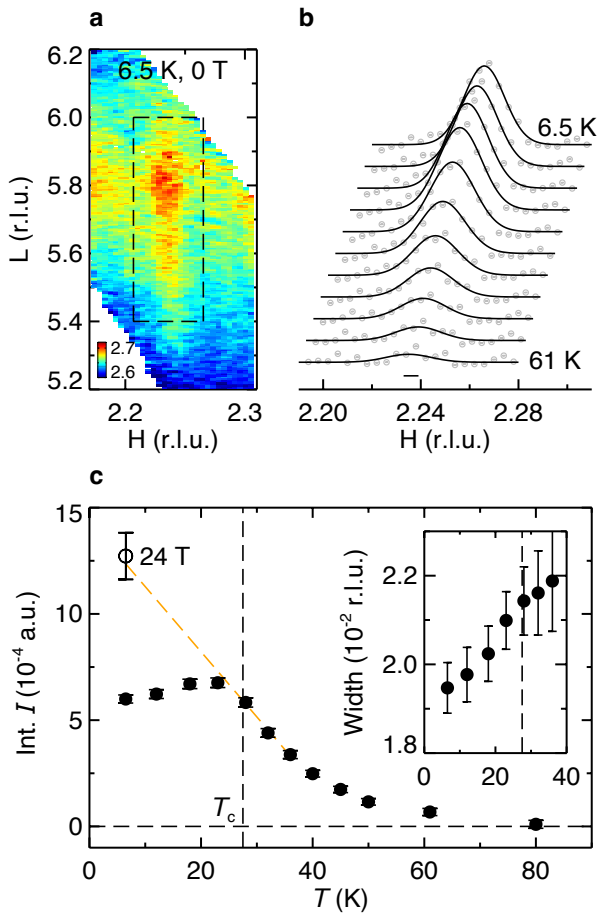}%
 \caption{\label{fig:1}Temperature dependence of CDW in LSCO. (a) CDW intensity map measured at 6.5 K, 0 T, projected onto HL plane, integrated over K [-0.02,0.02] r.l.u.. The dashed black rectangle encircles the CDW peak. (b) Temperature-dependent H-cuts through the CDW peak. Solid lines are single-peak fits to the data. A linear background has been subtracted, and data are shifted for clarity. The horizontal bar represents instrumental resolution. (c) Integrated CDW intensities extracted from the single-peak fits. The inset shows the corresponding peak width. The orange dashed line is a linear extrapolation of the CDW intensities at 28~K, 32~K, and 36~K to lower temperatures $T<T_c$, as described in the main text. The open circle shows the CDW intensity measured at 6.5 K, 24 T. Error bars represent one standard deviation.}
\end{figure}

 \begin{figure*}[t]
 \includegraphics[scale=0.6]{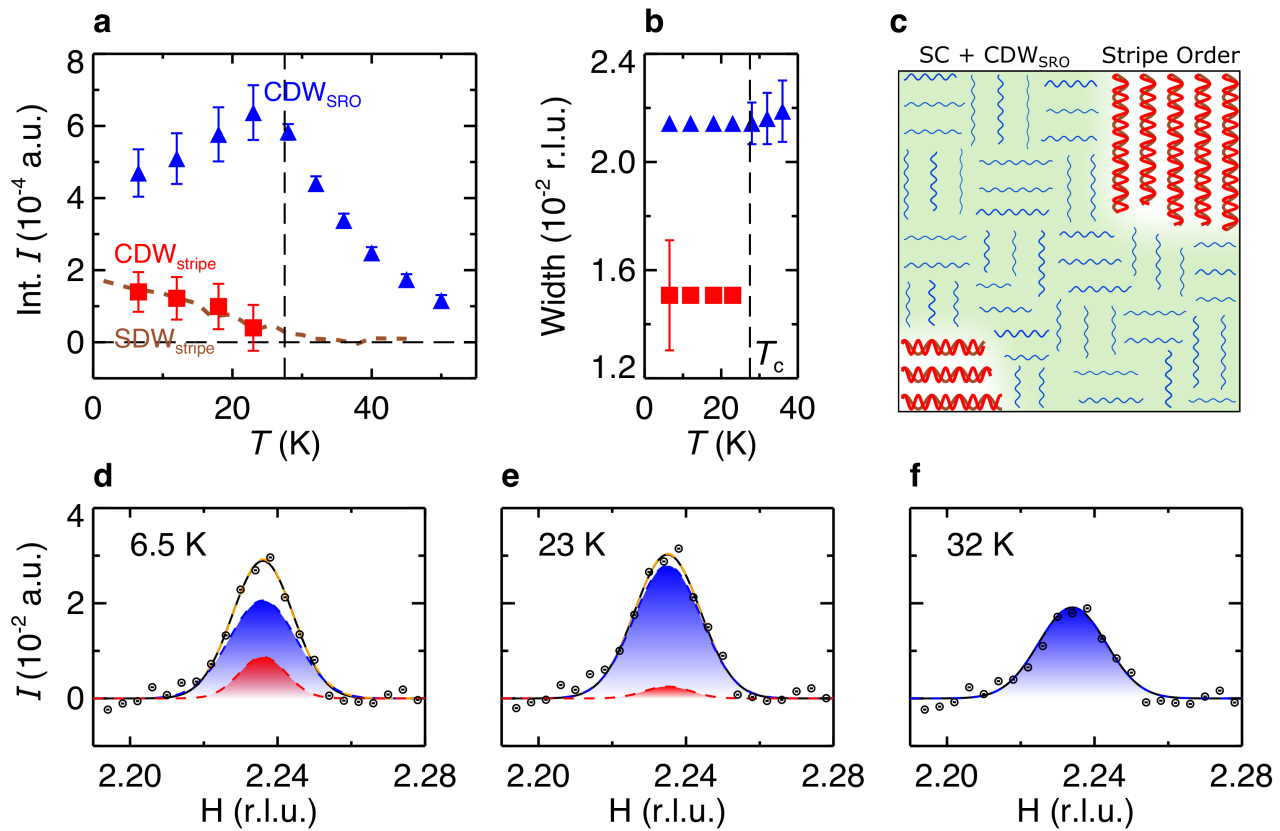}%
 \caption{\label{fig:2}Two-component decomposition of the CDW order in LSCO. (a) Integrated intensities for CDW$_{\rm SRO}$ and CDW$_{\rm stripe}$ extracted from the two-component fits. The brown dashed line shows the scaled SDW intensities measured by neutron scattering \cite{PhysRevB.59.6517}. (b) Corresponding peak width for CDW$_{\rm SRO}$ and CDW$_{\rm stripe}$. The widths are fixed for $T < T_c$ for the two-component fits, as described in the main text. (c) A pictorial illustration of the mixed phases in LSCO at low temperatures, which consist of superconducting regions with suppressed CDW$_{\rm SRO}$, and separate regions dominated by spin-charge stripe order. (d) - (f) Representative data and corresponding fits at 6.5 K, 23 K, and 32 K, respectively. Orange, blue, and red dashed lines show the total, CDW$_{\rm SRO}$, and CDW$_{\rm stripe}$ of the two-component fits, respectively. For comparison, the single-peak fits are shown in solid black lines. Error bars represent one standard deviation. }
 \end{figure*}

We first address the issue of inhomogeneity by examining the CDW $T$-dependence in zero magnetic field. As shown in Fig.~\ref{fig:1}a, the CDW peak appears as a rod of intensities along the L direction, which demonstrates its quasi-two-dimensional nature \cite{Ghiringhelli821,Chang2012,christensen2014,PhysRevB.89.224513} (more in Supplementary Figs.~5-6). To focus on the CDW correlations within the CuO$_2$ planes, the CDW intensities are integrated and projected along H. As a first analysis, a single Gaussian peak is found to fit the data well (Fig.~\ref{fig:1}b). Consistent with previous hard x-ray measurements \cite{PhysRevB.89.224513,christensen2014}, the CDW intensities become appreciable below $\sim$ 80 K (Fig.~\ref{fig:1}c). The onset with upward concavity is a common feature of the CDW in the cuprates, which indicates a lack of long-range CDW transition \cite{Hayward1336}. Resonant x-ray scattering has detected CDW in LSCO up to higher temperatures \cite{PhysRevB.90.100510,Wen2019}, which can be related to dynamical CDW fluctuations \cite{Arpaia906}. 

The CDW intensity reaches a maximum near the superconducting $T_c$ before getting suppressed at lower temperatures, indicating competition between CDW and superconductivity. However, unlike a homogeneously weakened order, the CDW peak width keeps decreasing for $T<T_c$ (Fig.~\ref{fig:1}c inset), which implies a growing CDW correlation length. Such contradictory behavior between the CDW intensity and correlation length has been observed, though with varying clarity, in previous x-ray measurements for LSCO at similar doping levels \cite{PhysRevB.89.224513,PhysRevB.90.100510,christensen2014}, and is in stark contrast with YBCO, where both the CDW intensity and correlation length decrease for $T<T_c$ \cite{Ghiringhelli821,Chang2012}. This indicates that mere competition between CDW and superconductivity is insufficient to explain the CDW behavior in LSCO. 

 \begin{figure*}[t]
 \includegraphics[scale=0.55]{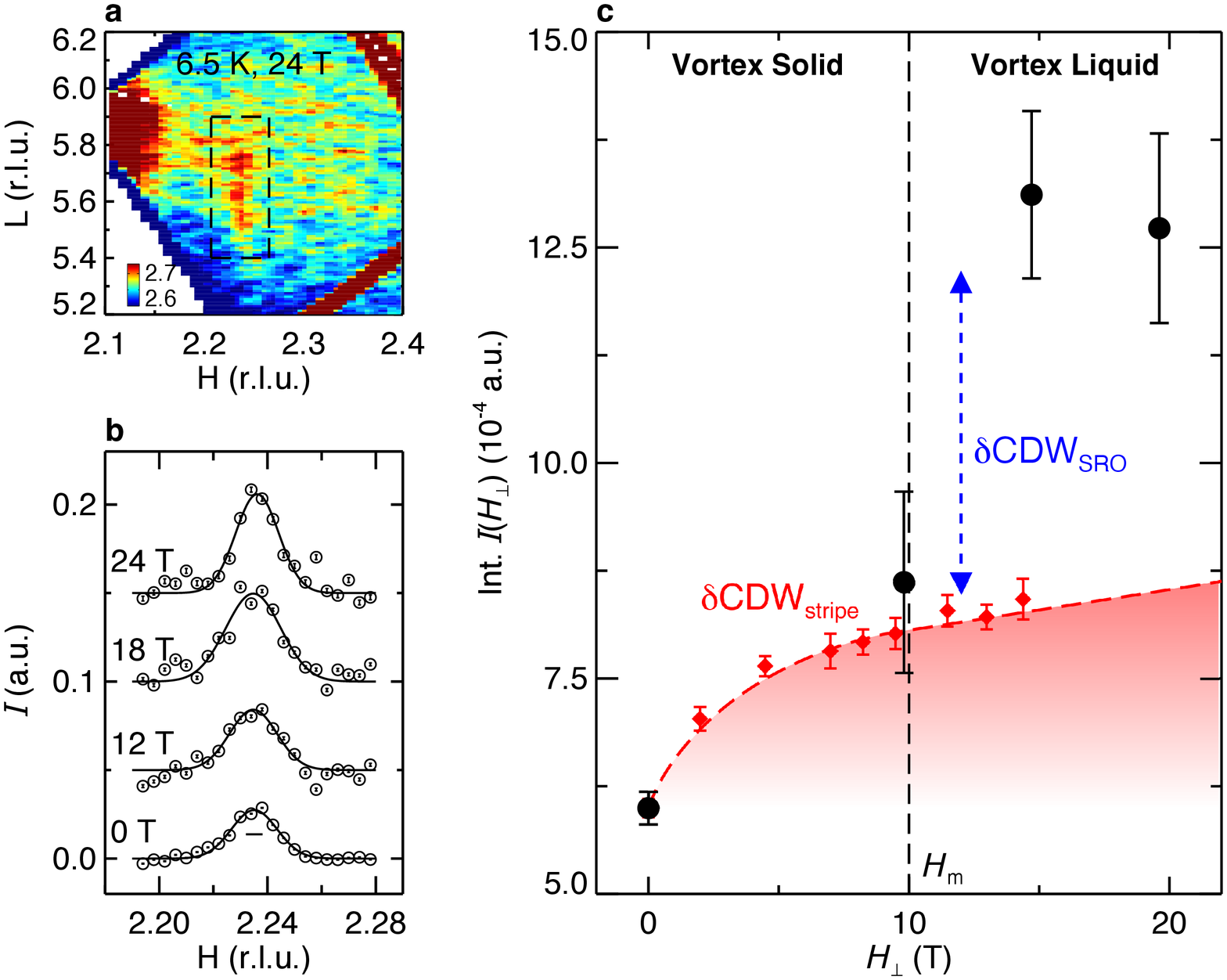}%
 \caption{\label{fig:3}Magnetic-field dependence of CDW in LSCO. (a) CDW intensity map measured at 6.5 K, 24 T, projected onto HL plane. (b) Magnetic-field-dependent H-cuts through the CDW peak. Solid lines are one-Gaussian fits to the data. A linear background has been subtracted, and data are shifted for clarity. The horizontal bar represents instrumental resolution. (c) Integrated CDW intensities as a function of magnetic field. Note here the magnetic-field projection perpendicular to the CuO$_2$ planes $H_\perp$ is plotted. The vertical dashed line marks the vortex-melting field $H_{\rm m}$ for LSCO at $T\sim 6.5$~K determined by in-plane resistivity measurements \cite{PhysRevB.103.115133}. The red diamonds are field-dependent SDW intensities measured by neutron for LSCO \cite{PhysRevB.78.104525}, scaled such to deduce the magnetic-field-induced enhancement of CDW$_{\rm stripe}$, as described in the main text. The red dashed line is guide to the eye. Error bars represent one standard deviation.}
 \end{figure*}

As alluded to above, the cause here is the SDW order which onsets at $T_{\rm sdw} \sim T_c$ in LSCO and enhances CDW \cite{PhysRevB.59.6517,Wen2019}, but does not coexist with CDW in YBCO \cite{RevModPhys.87.457}. To further elucidate the thermal evolution of the CDW, we decompose the CDW peak into two components: CDW$_{\rm stripe}$ to describe the SDW-enhanced component, and CDW$_{\rm SRO}$ (short-range order) to account for the competition with superconductivity. CDW$_{\rm SRO}$ is expected to behave similarly to the CDW in LSCO near optimal doping ($x\sim 0.145$) where the SDW order is absent \cite{Wen2019}. Since the CDW peak width there is weakly $T$-dependent at low temperatures \cite{Wen2019}, we fix the CDW$_{\rm SRO}$ peak width for $T<T_c$ to that extracted at $T_c$ in the single-peak fit, where the SDW order is just about to develop (Fig.~\ref{fig:2}b). This CDW$_{\rm SRO}$ component has correlation length of $\xi_{\rm SRO}=2/\mathrm{FWHM}=56(2)$~\AA. As for CDW$_{\rm stripe}$, to a first approximation we also assume a $T$-independent peak width. To extract this width, we fit the $T=6.5$~K data with the sum of the aforementioned CDW$_{\rm SRO}$ peak and a second peak. The best fit yields a sharper component (CDW$_{\rm stripe}$) with correlation length $\xi_{\rm stripe}=80(11)$~\AA. Fixing the widths for these two components and allowing the respective intensities to vary, such constrained two-component fitting provides an alternative and better account of the data (Figs.~\ref{fig:2}d-f, Supplementary Fig.~7). Consistent with our hypothesis, the fitting results show that the CDW $T$-dependence for $T<T_c$ can be described in terms of the weakening of CDW$_{\rm SRO}$ and the enhancement of CDW$_{\rm stripe}$ (Fig.~\ref{fig:2}a). In particular, the CDW$_{\rm stripe}$ intensity follows the SDW intensity measured by neutron scattering \cite{PhysRevB.59.6517} (Fig.~\ref{fig:2}a), consistent with their cooperative interactions. 

Such a decomposition provides a natural explanation of the seemingly contradictory behavior of the intensities and widths in the single-peak analysis. The combined effect of a weakened CDW$_{\rm SRO}$ (broad peak) and an enhanced CDW$_{\rm stripe}$ (sharp peak) results in the net reduction of the integrated intensity simultaneously with a narrowing of the width. Previous local probe measurements (such as $\mu$SR \cite{PhysRevB.66.014524} and NMR \cite{arsenault2019}) have indicated heterogeneous phases, but this has been difficult to directly corroborate with scattering studies which measure average correlations in the bulk. However, the high statistical quality of the zero-field data and resulting success of the two-component analysis to the CDW $T$-dependence strongly support the assertion that the CDW order is indeed heterogeneous for $T<T_c$ in LSCO, with coexisting CDW$_{\rm SRO}$ and CDW$_{\rm stripe}$ components. This further indicates superconductivity and SDW does not coexist in a uniform phase: the CuO$_2$ plane segregates into superconducting regions with suppressed CDW$_{\rm SRO}$, and separate regions dominated by spin-charge stripe order (CDW$_{\rm stripe}$ commensurate with SDW), as illustrated in Fig.~\ref{fig:2}c. 

With this picture in mind, we turn to the evolution of the CDW order under high magnetic field. Fig.~\ref{fig:3}a shows the intensity collected at $24$~T. The persistence of the rod-like scattering shows that the CDW correlations remain two-dimensional (more in Supplementary Fig.~5). We note that we find no evidence of three-dimensional CDW order at integer L position (here L$=6$) (see Supplementary Fig.~9), which has been observed in YBCO at a similar doping level and magnetic field \cite{Gerber949}. This difference could be related to the different value for $H_{c2}$ and/or how the CDW order on neighboring CuO$_2$ planes interacts \cite{PhysRevLett.119.107002}. Future measurements at even higher magnetic fields may provide more insight. 

Focusing on effects of field, we plot scans through the CDW peak in Fig.~\ref{fig:3}b. An increase in the CDW intensity with magnetic field is clearly observed. Although the statistical quality of the high-field data (constrained by magnet recovery time between magnetic pulses, see Methods) hinders a two-component analysis, the successful decomposition at zero field enables us to surmise the respective evolution of CDW$_{\rm SRO}$ and CDW$_{\rm stripe}$ in magnetic field. Assuming the proportionality between CDW$_{\rm stripe}$ and SDW (Fig.~\ref{fig:2}a) holds in magnetic field, we use the SDW magnetic-field dependence \cite{PhysRevB.78.104525} to infer the CDW$_{\rm stripe}$ enhancement (with no adjustable parameters), $\delta\mathrm{CDW}_{\mathrm{stripe}}(H_\perp)=\mathrm{CDW}_{\mathrm{stripe}}(0)\cdot[\mathrm{SDW}(H_\perp)/\mathrm{SDW}(0)-1]$, where $H_\perp$ is the applied field projection perpendicular to CuO$_2$ planes. As shown in Fig.~\ref{fig:3}c, $\delta\mathrm{CDW}_{\mathrm{stripe}}(H_\perp)$ can account for the overall CDW enhancement at $H_\perp \sim 10$~T. Comparison to prior CDW measurements \cite{christensen2014} shows $\delta\mathrm{CDW}_{\mathrm{stripe}}(H_\perp)$ also nicely describes the CDW evolution at smaller magnetic fields $H_\perp<10$~T (see Supplementary Fig.~8), suggesting that in the low field regime $<\sim 10$~T, the enhancement to the overall CDW is mainly due to CDW$_{\rm stripe}$, while the CDW$_{\rm SRO}$ intensity is relatively constant. Considering that CDW$_{\rm SRO}$ is expected to be strengthened as the magnetic field suppresses superconductivity \cite{Chang2012}, the volume fraction for CDW$_{\rm SRO}$ has likely been reduced by the field. This is consistent with $\mu$SR measurement that suggests the magnetic field increases the stripe order volume fraction \cite{PhysRevLett.95.157001}. 

At larger field $H_\perp > 10$~T, we observe a sudden increase in the overall CDW intensity that cannot be accounted for by $\delta\rm CDW_{\rm stripe}$ (Fig.~\ref{fig:3}c). We conclude this originates from CDW$_{\rm SRO}$. The field range where this enhancement occurs coincides with the vortex-melting field $H_m \sim$ 10~T inferred in transport measurements \cite{PhysRevB.73.024510,PhysRevB.103.115133}. Superconducting vortices which were pinned in the vortex solid at lower magnetic fields become mobile above $H_m$. This results in the loss of superconductivity due to the destruction of long-range phase coherence \cite{PhysRevB.73.024510}. The sudden increase of the CDW amplitude $\delta$CDW$_{\rm SRO}$ at high fields in Fig.~\ref{fig:3}c therefore implies a strong response of the CDW to mobile vortices. Such an intimate interaction between CDW and vortices may also be at play in recent reports of unusual nonohmic resistivity in the low-$T$ vortex-liquid state in cuprates \cite{Hsue2016275118}.

Both sets of our x-ray measurements (the $T$-dependence in zero-field, and the $H$-dependence at low-$T$) reveal the presence of two regions: regions that favor static spin-charge stripe order and regions that favor superconductivity (where the latter also harbor short-range CDW$_{\rm SRO}$). The stripe phase competes with uniform superconductivity, consistent with model calculations which reveal a near degeneracy between the superconducting and the stripe states \cite{Zheng1155}. Interestingly, a small change in the Hamiltonian can drive the system between these two distinct phases through an intermediate state featuring phase separation \cite{Jiang2020}. In LSCO, the presence of dopant disorder or structural inhomogeneity may play a role in stabilizing both phases simultaneously within the same sample.

The observed behavior of the CDW$_{\rm SRO}$ within the majority superconducting regions is particularly interesting. While CDW peaks have been observed in the normal and superconducting states of other cuprates, here, we find evidence that enhanced CDW$_{\rm SRO}$ is linked to the high-field vortex-liquid state. Noting that similar CDW peaks are observed in zero field in a broad range above $T_c$, one finds an interesting connection with Nernst effect measurements suggesting that the vortex-liquid state is continuously connected to the non-superconducting state above $T_c$, where both regions of the $H-T$ phase diagram possess a large Nernst signal due to mobile vortices \cite{PhysRevB.73.024510}. Our CDW measurements tell a complementary story. The CDW intensity in the vortex-liquid state (in a field of 24 T) is plotted in Fig.~\ref{fig:1}c, where we find good agreement with an extrapolation of the zero-field intensity from above $T_c$ to low temperatures. This connection between the CDW in the vortex liquid to that above $T_c$ indicates that mobile vortices and CDW$_{\rm SRO}$ correlations are both inherent to the state where superconducting long-range phase coherence is lost. Within a phase-disordering scenario for the loss of superconductivity \cite{PhysRevB.73.024510}, short-range two-dimensional CDW correlations appear to be compatible with local superconducting pairing.

These results reveal a distinction between the two types of charge order in the cuprates: CDW$_{\rm stripe}$ and CDW$_{\rm SRO}$. CDW$_{\rm stripe}$ and the associated static stripe state is most prominent in the La$_2$CuO$_4$-based cuprates, and is clearly competitive with uniform superconductivity. The ubiquitous CDW$_{\rm SRO}$, on the other hand, coexists with local superconductivity and may even aid the formation of vortices \cite{Sachdev2003,PhysRevLett.92.177002}. It is long range superconducting phase coherence that simultaneously suppresses CDW$_{\rm SRO}$ and the presence of mobile vortices. The apparent sensitivity of CDW correlations to superconducting phase coherence further suggests a unified quantum description of the density waves and superconductivity in cuprate superconductors \cite{tranquada_review,RevModPhys.87.457,Hayward1336}.\\

\noindent{{\bf Methods}}\\
Single crystalline La$_{1.885}$Sr$_{0.115}$CuO$_4$ samples were grown by the traveling solvent floating zone method. The typical growth rate was 1.0 mm h$^{-1}$ and a 50 - 60 mm-long crystal rod was successfully obtained. A 10 mm-long crystalline piece from the end part of the grown rod was annealed in oxygen gas flow to minimize oxygen deficiencies. The superconducting transition temperature $T_c$ of the sample is determined to be 27.5(2)~K (Supplementary Fig.~2). We focus on the CDW peak near (2.23,0,5.5) r.l.u. (reciprocal space is denoted using the tetragonal unit cell, a = b = 3.77~\AA, c = 13.25~\AA), where the CDW intensity is strong \cite{PhysRevB.89.224513} and the scattering geometry allows a large magnetic field projection ($\sim82\%$) along the crystallographic $c$-axis (perpendicular to the CuO$_2$ planes). The CDW peaks were initially confirmed by resonant soft x-ray scattering at SSRL \cite{Wen2019}. A $1\times0.5\times0.5$~mm$^3$ sample was oriented using Laue x-ray diffraction and polished to the desired dimensions with [2.23,0,5.5] direction normal to the scattering surface. The vertical scattering plane is spanned by the nominal [2.23,0,5.5] and [0,1,0] directions. In this geometry, the magnetic field direction is $\sim35.2^\circ$ tilted away from the crystalline $c$-axis.  

The x-ray scattering experiment was carried out on the X-ray Correlation Spectroscopy (XCS) instrument at the Linac Coherent Light Source (LCLS) at the SLAC National Accelerator Laboratory (Supplementary Fig.~1). The orientation of the sample was determined by measuring the (204) and (206) nuclear Bragg peaks, which was then used to convert the pixel coordinate to the reciprocal-space coordinate (Supplementary Fig.~3). The CDW temperature dependence was probed upon warming. The magnetic-field-dependent measurements were carried out at the lowest temperature achievable of 6.5~K, where the largest field-induced effect is expected. The sample rotation was fixed at the CDW rocking scan peak center (Supplementary Fig.~4). For each magnetic-field run, 10 measurements at zero field were taken immediately before and after the field pulse to provide an accurate zero-field reference (Supplementary Figure.~1). While for zero-field measurements data can be taken continuously at 120 Hz x-ray pulse frequency, measurements in magnetic fields require extra time to cool down the magnet after each magnetic-field pulse. For example, at 24~T the cool-down time was typically $\sim 15$ minutes between magnetic pulses. This imposed a constraint on the statistical quality of the high-field data achievable during the finite beam time.\\ 

\noindent{{\bf Data availability}}\\
All data that support the plots and other findings of this study are available from the corresponding authors upon reasonable request.\\ 

\noindent{{\bf Acknowledgements}}\\
We acknowledge W.-S. Lee, S. A. Kivelson, T. P. Devereaux for insightful discussions. This work is supported by the U.S. Department of Energy (DOE), Office of Science, Basic Energy Sciences, Materials Sciences and Engineering Division, under contract DE-AC02-76SF00515. X-ray FEL studies were carried out at the Linac Coherent Light Source, a Directorate of SLAC and an Office of Science User Facility operated for the DOE, Office of Science by Stanford University. Soft X-ray characterization measurements were carried out at the Stanford Synchrotron Radiation Lightsource (beamline 13–3), SLAC National Accelerator Laboratory, supported by the U.S. Department of Energy, Office of Science, Office of Basic Energy Sciences under Contract No. DE-AC02-76SF00515. H.N. acknowledges the support by Grants-in-Aid for Scientific Research (KAKENHI) 23224009, International Collaboration Center-Institute for Materials Research, and MD-program. M.F. is supported by JSPS KAKENHI under Grants Nos. 16H02125 and 21H04987. H.J. acknowledges the support by the National Research Foundation grant funded by the Korea government (MSIT) (Grant No. 2019R1F1A1060295). Part of this work was performed at the Stanford Nano Shared Facilities (SNSF), supported by the National Science Foundation under award ECCS-2026822.\\

\noindent{{\bf Author contributions}}\\
J.W., H.J., H.N., C.K., J.L. and Y.L. designed the study. J.W., W.H., H.J., H.N., S.M., S.S., M.C., D.Z., Y.L., J.M.J., C.R.R., J.L., and Y.L. carried out the experiment. H.N and S.M supported the operation of the pulsed magnet. M.F. synthesized the sample. J.W. analyzed the data. J.W., H.C.J., J.L., and Y.L. wrote the manuscript with critical inputs from all authors. \\

\noindent{{\bf Competing interests}}\\
The authors declare no competing interests.


\begin{thebibliography}{10}
\expandafter\ifx\csname url\endcsname\relax
  \def\url#1{\texttt{#1}}\fi
\expandafter\ifx\csname urlprefix\endcsname\relax\def\urlprefix{URL }\fi
\providecommand{\bibinfo}[2]{#2}
\providecommand{\eprint}[2][]{\url{#2}}

\bibitem{Keimer2015}
\bibinfo{author}{Keimer, B.}, \bibinfo{author}{Kivelson, S.~A.},
  \bibinfo{author}{Norman, M.~R.}, \bibinfo{author}{Uchida, S.} \&
  \bibinfo{author}{Zaanen, J.}
\newblock \bibinfo{title}{From quantum matter to high-temperature
  superconductivity in copper oxides}.
\newblock \emph{\bibinfo{journal}{Nature}} \textbf{\bibinfo{volume}{518}},
  \bibinfo{pages}{179--186} (\bibinfo{year}{2015}).

\bibitem{tranquada_review}
\bibinfo{author}{Tranquada, J.~M.}
\newblock \bibinfo{title}{Cuprate superconductors as viewed through a striped
  lens}.
\newblock \emph{\bibinfo{journal}{Adv. Phys.}} \textbf{\bibinfo{volume}{69}},
  \bibinfo{pages}{437--509} (\bibinfo{year}{2020}).

\bibitem{RevModPhys.87.457}
\bibinfo{author}{Fradkin, E.}, \bibinfo{author}{Kivelson, S.~A.} \&
  \bibinfo{author}{Tranquada, J.~M.}
\newblock \bibinfo{title}{Colloquium: Theory of intertwined orders in high
  temperature superconductors}.
\newblock \emph{\bibinfo{journal}{Rev. Mod. Phys.}}
  \textbf{\bibinfo{volume}{87}}, \bibinfo{pages}{457--482}
  (\bibinfo{year}{2015}).

\bibitem{Dagotto257}
\bibinfo{author}{Dagotto, E.}
\newblock \bibinfo{title}{Complexity in strongly correlated electronic
  systems}.
\newblock \emph{\bibinfo{journal}{Science}} \textbf{\bibinfo{volume}{309}},
  \bibinfo{pages}{257--262} (\bibinfo{year}{2005}).

\bibitem{EMERY1993597}
\bibinfo{author}{Emery, V.~J.} \& \bibinfo{author}{Kivelson, S.~A.}
\newblock \bibinfo{title}{Frustrated electronic phase separation and
  high-temperature superconductors}.
\newblock \emph{\bibinfo{journal}{Physica C: Superconductivity}}
  \textbf{\bibinfo{volume}{209}}, \bibinfo{pages}{597--621}
  (\bibinfo{year}{1993}).

\bibitem{MARTIN200146}
\bibinfo{author}{Martin, I.} \& \bibinfo{author}{Balatsky, A.~V.}
\newblock \bibinfo{title}{Doping-induced inhomogeneity in {high-$T_c$}
  superconductors}.
\newblock \emph{\bibinfo{journal}{Physica C: Superconductivity}}
  \textbf{\bibinfo{volume}{357-360}}, \bibinfo{pages}{46--48}
  (\bibinfo{year}{2001}).

\bibitem{pan2001}
\bibinfo{author}{Pan, S.~H.} \emph{et~al.}
\newblock \bibinfo{title}{Microscopic electronic inhomogeneity in the
  high-{$T_c$} superconductor {Bi$_2$Sr$_2$CaCu$_2$O$_{8+x}$}}.
\newblock \emph{\bibinfo{journal}{Nature}} \textbf{\bibinfo{volume}{413}},
  \bibinfo{pages}{282--285} (\bibinfo{year}{2001}).

\bibitem{KRESIN2006231}
\bibinfo{author}{Kresin, V.~Z.}, \bibinfo{author}{Ovchinnikov, Y.~N.} \&
  \bibinfo{author}{Wolf, S.~A.}
\newblock \bibinfo{title}{Inhomogeneous superconductivity and the
  “pseudogap” state of novel superconductors}.
\newblock \emph{\bibinfo{journal}{Phys. Rep.}} \textbf{\bibinfo{volume}{431}},
  \bibinfo{pages}{231--259} (\bibinfo{year}{2006}).

\bibitem{Zhang1089}
\bibinfo{author}{Zhang, S.-C.}
\newblock \bibinfo{title}{A unified theory based on $\mathit{SO}$(5) symmetry
  of superconductivity and antiferromagnetism}.
\newblock \emph{\bibinfo{journal}{Science}} \textbf{\bibinfo{volume}{275}},
  \bibinfo{pages}{1089--1096} (\bibinfo{year}{1997}).

\bibitem{Kivelson11903}
\bibinfo{author}{Kivelson, S.~A.}, \bibinfo{author}{Aeppli, G.} \&
  \bibinfo{author}{Emery, V.~J.}
\newblock \bibinfo{title}{Thermodynamics of the interplay between magnetism and
  high-temperature superconductivity}.
\newblock \emph{\bibinfo{journal}{Proc. Nat. Acad. Sci. USA}}
  \textbf{\bibinfo{volume}{98}}, \bibinfo{pages}{11903--11907}
  (\bibinfo{year}{2001}).

\bibitem{PhysRevLett.87.067202}
\bibinfo{author}{Demler, E.}, \bibinfo{author}{Sachdev, S.} \&
  \bibinfo{author}{Zhang, Y.}
\newblock \bibinfo{title}{Spin-ordering quantum transitions of superconductors
  in a magnetic field}.
\newblock \emph{\bibinfo{journal}{Phys. Rev. Lett.}}
  \textbf{\bibinfo{volume}{87}}, \bibinfo{pages}{067202}
  (\bibinfo{year}{2001}).

\bibitem{PhysRevB.66.014524}
\bibinfo{author}{Savici, A.~T.} \emph{et~al.}
\newblock \bibinfo{title}{Muon spin relaxation studies of incommensurate
  magnetism and superconductivity in stage-4
  {${\mathrm{La}}_{2}{\mathrm{CuO}}_{4.11}$ and
  ${\mathrm{La}}_{1.88}{\mathrm{Sr}}_{0.12}{\mathrm{CuO}}_{4}$}}.
\newblock \emph{\bibinfo{journal}{Phys. Rev. B}} \textbf{\bibinfo{volume}{66}},
  \bibinfo{pages}{014524} (\bibinfo{year}{2002}).

\bibitem{Mohottala2006}
\bibinfo{author}{Mohottala, H.~E.} \emph{et~al.}
\newblock \bibinfo{title}{Phase separation in superoxygenated
  $\mathrm{La}_{2-x}\mathrm{Sr}_{x}\mathrm{CuO}_{4+y}$}.
\newblock \emph{\bibinfo{journal}{Nat. Mater.}} \textbf{\bibinfo{volume}{5}},
  \bibinfo{pages}{377--382} (\bibinfo{year}{2006}).

\bibitem{PhysRevB.78.104525}
\bibinfo{author}{Chang, J.} \emph{et~al.}
\newblock \bibinfo{title}{{Tuning competing orders in
  ${\text{La}}_{2\ensuremath{-}x}{\text{Sr}}_{x}{\text{CuO}}_{4}$ cuprate
  superconductors by the application of an external magnetic field}}.
\newblock \emph{\bibinfo{journal}{Phys. Rev. B}} \textbf{\bibinfo{volume}{78}},
  \bibinfo{pages}{104525} (\bibinfo{year}{2008}).

\bibitem{Ghiringhelli821}
\bibinfo{author}{Ghiringhelli, G.} \emph{et~al.}
\newblock \bibinfo{title}{Long-range incommensurate charge fluctuations in
  {(Y,Nd)Ba$_2$Cu$_3$O$_{6+x}$}}.
\newblock \emph{\bibinfo{journal}{Science}} \textbf{\bibinfo{volume}{337}},
  \bibinfo{pages}{821--825} (\bibinfo{year}{2012}).

\bibitem{Chang2012}
\bibinfo{author}{Chang, J.} \emph{et~al.}
\newblock \bibinfo{title}{Direct observation of competition between
  superconductivity and charge density wave order in
  {YBa$_2$Cu$_3$O$_{6.67}$}}.
\newblock \emph{\bibinfo{journal}{Nat. Phys.}} \textbf{\bibinfo{volume}{8}},
  \bibinfo{pages}{871--876} (\bibinfo{year}{2012}).

\bibitem{PhysRevB.57.1422}
\bibinfo{author}{Zachar, O.}, \bibinfo{author}{Kivelson, S.~A.} \&
  \bibinfo{author}{Emery, V.~J.}
\newblock \bibinfo{title}{Landau theory of stripe phases in cuprates and
  nickelates}.
\newblock \emph{\bibinfo{journal}{Phys. Rev. B}} \textbf{\bibinfo{volume}{57}},
  \bibinfo{pages}{1422--1426} (\bibinfo{year}{1998}).

\bibitem{Wen2019}
\bibinfo{author}{Wen, J.-J.} \emph{et~al.}
\newblock \bibinfo{title}{Observation of two types of charge-density-wave
  orders in superconducting {La$_{2-x}$Sr$_x$CuO$_4$}}.
\newblock \emph{\bibinfo{journal}{Nat. Commun.}} \textbf{\bibinfo{volume}{10}},
  \bibinfo{pages}{3269} (\bibinfo{year}{2019}).

\bibitem{PhysRevLett.126.167001}
\bibinfo{author}{Huang, H.} \emph{et~al.}
\newblock \bibinfo{title}{Two-dimensional superconducting fluctuations
  associated with charge-density-wave stripes in
  $\mathrm{La}_{1.87}\mathrm{Sr}_{0.13}\mathrm{Cu}_{0.99}\mathrm{Fe}_{0.01}\mathrm{O}_{4}$}.
\newblock \emph{\bibinfo{journal}{Phys. Rev. Lett.}}
  \textbf{\bibinfo{volume}{126}}, \bibinfo{pages}{167001}
  (\bibinfo{year}{2021}).

\bibitem{PhysRevB.89.224513}
\bibinfo{author}{Croft, T.~P.}, \bibinfo{author}{Lester, C.},
  \bibinfo{author}{Senn, M.~S.}, \bibinfo{author}{Bombardi, A.} \&
  \bibinfo{author}{Hayden, S.~M.}
\newblock \bibinfo{title}{Charge density wave fluctuations in
  {${\text{La}}_{2\ensuremath{-}x}$${\text{Sr}}_{x}$${\text{CuO}}_{4}$} and
  their competition with superconductivity}.
\newblock \emph{\bibinfo{journal}{Phys. Rev. B}} \textbf{\bibinfo{volume}{89}},
  \bibinfo{pages}{224513} (\bibinfo{year}{2014}).

\bibitem{PhysRevB.90.100510}
\bibinfo{author}{Thampy, V.} \emph{et~al.}
\newblock \bibinfo{title}{Rotated stripe order and its competition with
  superconductivity in
  {${\mathrm{La}}_{1.88}{\mathrm{Sr}}_{0.12}{\mathrm{CuO}}_{4}$}}.
\newblock \emph{\bibinfo{journal}{Phys. Rev. B}} \textbf{\bibinfo{volume}{90}},
  \bibinfo{pages}{100510(R)} (\bibinfo{year}{2014}).

\bibitem{christensen2014}
\bibinfo{author}{{Christensen}, N.~B.} \emph{et~al.}
\newblock \bibinfo{title}{Bulk charge stripe order competing with
  superconductivity in {La$_{2-x}$Sr$_x$CuO$_4$ (x=0.12)}}
  (\bibinfo{year}{2014}).
\newblock \urlprefix\url{https://arxiv.org/abs/1404.3192}.

\bibitem{PhysRevB.59.6517}
\bibinfo{author}{Kimura, H.} \emph{et~al.}
\newblock \bibinfo{title}{Neutron-scattering study of static antiferromagnetic
  correlations in
  {${\mathrm{La}}_{2\ensuremath{-}x}{\mathrm{Sr}}_{x}{\mathrm{Cu}}_{1\ensuremath{-}y}{\mathrm{Zn}}_{y}{\mathrm{O}}_{4}$}}.
\newblock \emph{\bibinfo{journal}{Phys. Rev. B}} \textbf{\bibinfo{volume}{59}},
  \bibinfo{pages}{6517--6523} (\bibinfo{year}{1999}).

\bibitem{Hayward1336}
\bibinfo{author}{Hayward, L.~E.}, \bibinfo{author}{Hawthorn, D.~G.},
  \bibinfo{author}{Melko, R.~G.} \& \bibinfo{author}{Sachdev, S.}
\newblock \bibinfo{title}{Angular fluctuations of a multicomponent order
  describe the pseudogap of {YBa$_2$Cu$_3$O$_{6+x}$}}.
\newblock \emph{\bibinfo{journal}{Science}} \textbf{\bibinfo{volume}{343}},
  \bibinfo{pages}{1336--1339} (\bibinfo{year}{2014}).

\bibitem{Arpaia906}
\bibinfo{author}{Arpaia, R.} \emph{et~al.}
\newblock \bibinfo{title}{Dynamical charge density fluctuations pervading the
  phase diagram of a {Cu}-based high-{$T_c$} superconductor}.
\newblock \emph{\bibinfo{journal}{Science}} \textbf{\bibinfo{volume}{365}},
  \bibinfo{pages}{906--910} (\bibinfo{year}{2019}).

\bibitem{PhysRevB.103.115133}
\bibinfo{author}{Frachet, M.} \emph{et~al.}
\newblock \bibinfo{title}{{High magnetic field ultrasound study of spin
  freezing in ${\mathrm{La}}_{1.88}{\mathrm{Sr}}_{0.12}{\mathrm{CuO}}_{4}$}}.
\newblock \emph{\bibinfo{journal}{Phys. Rev. B}}
  \textbf{\bibinfo{volume}{103}}, \bibinfo{pages}{115133}
  (\bibinfo{year}{2021}).

\bibitem{arsenault2019}
\bibinfo{author}{Arsenault, A.}, \bibinfo{author}{Imai, T.},
  \bibinfo{author}{Singer, P.~M.}, \bibinfo{author}{Suzuki, K.~M.} \&
  \bibinfo{author}{Fujita, M.}
\newblock \bibinfo{title}{Magnetic inhomogeneity in charge-ordered
  {${\mathrm{La}}_{1.885}{\mathrm{Sr}}_{0.115}{\mathrm{CuO}}_{4}$} studied by
  {NMR}}.
\newblock \emph{\bibinfo{journal}{Phys. Rev. B}}
  \textbf{\bibinfo{volume}{101}}, \bibinfo{pages}{184505}
  (\bibinfo{year}{2020}).

\bibitem{Gerber949}
\bibinfo{author}{Gerber, S.} \emph{et~al.}
\newblock \bibinfo{title}{Three-dimensional charge density wave order in
  {YBa$_2$Cu$_3$O$_{6.67}$} at high magnetic fields}.
\newblock \emph{\bibinfo{journal}{Science}} \textbf{\bibinfo{volume}{350}},
  \bibinfo{pages}{949--952} (\bibinfo{year}{2015}).

\bibitem{PhysRevLett.119.107002}
\bibinfo{author}{Caplan, Y.} \& \bibinfo{author}{Orgad, D.}
\newblock \bibinfo{title}{Dimensional crossover of charge-density wave
  correlations in the cuprates}.
\newblock \emph{\bibinfo{journal}{Phys. Rev. Lett.}}
  \textbf{\bibinfo{volume}{119}}, \bibinfo{pages}{107002}
  (\bibinfo{year}{2017}).

\bibitem{PhysRevLett.95.157001}
\bibinfo{author}{Savici, A.~T.} \emph{et~al.}
\newblock \bibinfo{title}{Muon spin relaxation studies of
  magnetic-field-induced effects in high-${T}_{c}$ superconductors}.
\newblock \emph{\bibinfo{journal}{Phys. Rev. Lett.}}
  \textbf{\bibinfo{volume}{95}}, \bibinfo{pages}{157001}
  (\bibinfo{year}{2005}).

\bibitem{PhysRevB.73.024510}
\bibinfo{author}{Wang, Y.}, \bibinfo{author}{Li, L.} \& \bibinfo{author}{Ong,
  N.~P.}
\newblock \bibinfo{title}{Nernst effect in high-{${T}_{c}$} superconductors}.
\newblock \emph{\bibinfo{journal}{Phys. Rev. B}} \textbf{\bibinfo{volume}{73}},
  \bibinfo{pages}{024510} (\bibinfo{year}{2006}).

\bibitem{Hsue2016275118}
\bibinfo{author}{Hsu, Y.-T.} \emph{et~al.}
\newblock \bibinfo{title}{Anomalous vortex liquid in charge-ordered cuprate
  superconductors}.
\newblock \emph{\bibinfo{journal}{Proc. Nat. Acad. Sci. USA}}
  \textbf{\bibinfo{volume}{118}}, \bibinfo{pages}{e2016275118}
  (\bibinfo{year}{2021}).

\bibitem{Zheng1155}
\bibinfo{author}{Zheng, B.-X.} \emph{et~al.}
\newblock \bibinfo{title}{{Stripe order in the underdoped region of the
  two-dimensional Hubbard model}}.
\newblock \emph{\bibinfo{journal}{Science}} \textbf{\bibinfo{volume}{358}},
  \bibinfo{pages}{1155--1160} (\bibinfo{year}{2017}).

\bibitem{Jiang2020}
\bibinfo{author}{Jiang, Y.-F.}, \bibinfo{author}{Zaanen, J.},
  \bibinfo{author}{Devereaux, T.~P.} \& \bibinfo{author}{Jiang, H.-C.}
\newblock \bibinfo{title}{{Ground state phase diagram of the doped Hubbard
  model on the four-leg cylinder}}.
\newblock \emph{\bibinfo{journal}{Phys. Rev. Research}}
  \textbf{\bibinfo{volume}{2}}, \bibinfo{pages}{033073} (\bibinfo{year}{2020}).

\bibitem{Sachdev2003}
\bibinfo{author}{Sachdev, S.}
\newblock \emph{\bibinfo{title}{Spin and Charge Order in The Vortex Lattice of
  The Cuprates: Experiment and Theory}}, \bibinfo{pages}{171--186}
  (\bibinfo{publisher}{Springer US}, \bibinfo{address}{Boston, MA},
  \bibinfo{year}{2003}).

\bibitem{PhysRevLett.92.177002}
\bibinfo{author}{Honerkamp, C.} \& \bibinfo{author}{Lee, P.~A.}
\newblock \bibinfo{title}{Staggered flux vortices and the superconducting
  transition in the layered cuprates}.
\newblock \emph{\bibinfo{journal}{Phys. Rev. Lett.}}
  \textbf{\bibinfo{volume}{92}}, \bibinfo{pages}{177002}
  (\bibinfo{year}{2004}).

\end{thebibliography}
\end{document}